\documentclass[aps,prd,preprintnumbers,twocolumn,floatfix,nofootinbib,showpacs]{revtex4}

\preprint{MADPH-09-1533}

\usepackage{graphicx}
\usepackage{ulem}
\usepackage{amsmath,amssymb}
\usepackage{url}
\usepackage{color}
\usepackage{multirow}

\newcommand{\lsim}{\mathrel{\mathop{\kern 0pt \rlap
  {\raise.2ex\hbox{$<$}}}
  \lower.9ex\hbox{\kern-.190em $\sim$}}}
\newcommand{\gsim}{\mathrel{\mathop{\kern 0pt \rlap
  {\raise.2ex\hbox{$>$}}}
  \lower.9ex\hbox{\kern-.190em $\sim$}}}

\newcommand{\gev}{{\,{\rm GeV}}}
\newcommand{\tev}{{\,{\rm TeV}}}
\newcommand{\gm}{{\gamma}}

\newcommand{\Gm}{{\Gamma}}

\newcommand{\beq}{\begin{equation}}
\newcommand{\eeq}{\end{equation}}
\newcommand{\bea}{\begin{eqnarray}}
\newcommand{\eea}{\end{eqnarray}}
\newcommand{\no}{{\nonumber}}

\newcommand{\et}{{\rlap/\!E_T}}

\newcommand{\Lg}{{\mathcal{L}}}

\newcommand{\zt}{ { Z^{(2)} }}
\newcommand{\lo}{ { L^{(1)} }}

\newcommand{\gmo}{ {\gamma^{(1)}}}

\newcommand{\neu}{\tilde{\chi}^0}

\newcommand{\neuo}{{\tilde{\chi}^0_1}}

\newcommand{\slt}{{\tilde{\ell}}}

\newcommand{\mb} {m_B^{}}

\begin{document}

\title{Kinematic Cusps: Determining the Missing Particle Mass at Colliders}
\bigskip
\author{Tao Han$^{1}$}
\author{Ian-Woo Kim$^1$}
\author{Jeonghyeon Song$^{2}$}
\affiliation{$^1$Department of Physics, University of Wisconsin, Madison, WI 53706, USA}
\affiliation{$^2$Division of Quantum Phases \& Devices,
School of Physics, Konkuk University, Seoul 143-701, Korea}
\date{\today}
\begin{abstract}
In many extensions of the SM, neutral massive stable particles 
(dark matter candidates) are produced
at colliders in pairs  due to an exact symmetry called a ``parity''.
These particles escape detection, rendering their mass measurement difficult.
In the pair production of such particles via a specific (``antler'') decay
topology, kinematic cusp structures are present in the invariant mass and
angular distributions of the observable particles. Together with the 
end-points, such cusps can be used to measure the missing particle mass
and the intermediate particle mass in the decay chain.  
Our simulation of a benchmark scenario in a $Z'$ supersymmetric
model shows that the cusp feature
survives under the consideration of detector simulation
and the standard model backgrounds.
This technique for determining missing particle masses should be invaluable in the search
for new physics at the LHC and future lepton colliders. 
\end{abstract}
\maketitle


\section{Introduction}

Pauli's postulation of a new particle that escapes from detection and carries
away energy and angular momentum in $\beta$ decay not only laid out the foundation
for the weak interaction, but also rightfully
introduced the first dark matter  particle, the neutrino.
Ever since then, attempts to determine the masses and other properties  of  the neutrinos
have led to many research efforts in nuclear physics,
particle physics, astroparticle physics and cosmology.
If the upcoming experiments at the CERN LHC find evidence of large missing energy events
beyond the Standard Model (SM) expectations, 
this exciting discovery may 
hold the key to explain the missing mass puzzle
in the Universe,
the dark matter. It is thus of fundamental importance to determine the
mass and properties of
this missing particle in LHC experiments, to uncover its underlying dynamics
and to check its consistency with dark matter expectations.

This task is challenging, however, 
even with the establishment of missing energy events at the LHC.
In hadronic collisions, the undetermined longitudinal motion of the
parton-level scattering leads to the ambiguity of their c.m.~frame and thus 
the partonic c.m.~energy.
Furthermore, with a conserved discrete quantum number (generically called a ``parity'')
that keeps the lightest particle in the new sector stable,
the missing particles always come in pairs. 
The final state kinematics is even less constrained.

Great efforts have been made to reconstruct the mass of the missing particle. 
It is well known that in cascade decays, the masses of 
the invisible particles can be extracted from the maximum end-points of 
kinematic variables such as invariant mass distributions 
of leptons and 
jets~\cite{endpoint}.
Another interesting approach is to determine these parameters from 
transverse mass variables that utilize missing transverse energy, 
such as $M_{T2}$ for processes with a pair of missing particles~\cite{MT2}. 
The end-point of the $M_{T2}$ distribution is known to display a kink 
when the trial mass for the missing particle
is identical to the true 
mass~\cite{MT2kink}.
For long cascade decay chains, it is possible to
determine the unknown masses 
through the event reconstruction from 
the mass shell conditions by combining the
information from multiple events~\cite{mass:shell:cascade}.

\begin{figure}[tb]
\centering
  \includegraphics[width=6cm]{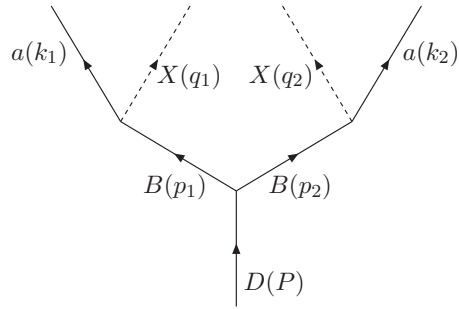}
  \caption{\label{fig:topology}The ``antler'' decay diagram, a heavy particle ($D$) to
two visible SM particles ($a$) and two missing particles ($X$), via two on-shell intermediate
particles ($B$).}
\end{figure}

 In this paper, we propose an additional method for the missing particle mass measurements, based on what we call ``kinematic
cusps", 
non-smooth substructures 
in kinematic distributions. 
Kinematic cusps arise in many new physics processes. We focus here on a particular class of processes that we dub as ``antler'' decays: the two missing parity-odd particles ($X$) come along with two visible SM
particles ($a$), from the decay of a heavy parity-even particle
($D$) through intermediate parity-odd particles ($B$), as shown in
Fig.~\ref{fig:topology}. 
The advantage of considering the kinematic cusps 
and end-points is that once the 
parent mass ($m_D$) is known, the masses of the missing particle 
($m_X$) and the intermediate particle ($\mb$) can be determined by 
measuring the energy-momenta of the visible particles 
without 
combinatoric complications.
Even though the kinematic cusps can be found in other processes,
we focus on this case due to its simplicity. 
Other examples will be 
presented elsewhere~\cite{workinprogress}.\footnote{General mass measurement 
techniques using such kinematic singularities have been recently 
developed~\cite{Kim:2009si}. }

The antler topology is common in many scenarios
with dark matter particle candidates. 
Familiar examples can be found in the following
theoretically well-motivated models:
\bea
\label{eq:model}
\hbox{MSSM~\cite{mssm},} && H \to \neu_2+ \neu_2 \to Z\neuo+ Z\neuo;
\nonumber \\
\hbox{$Z'$ SUSY~\cite{zpsusy},} && Z' \to \slt^- +\slt^+ \to \ell^-\neuo+\ell^+\neuo ;
\\
\hbox{UED~\cite{ued},} && \zt \to \lo +\lo  \to \ell^- \gmo+ \ell^+ \gmo ;
\nonumber
\\
\hbox{LHT~\cite{LHT},} && H \to t_- + t_-  \to t A_H + t A_H.
\nonumber
\eea
The precondition is that the mass of particle $D$ is known a priori.
This can be achieved
since its even-parity allows its decay into two observable SM particles.
In this regard, the antler topology
is equally applicable to a lepton collider 
where the c.m.~energy is accurately known.
Among these examples above, the decay of $Z'$ 
in a supersymmetric theory 
was studied  \cite{zpsusy} to 
measure the missing particle mass based on the $M_{T2}$ variable, 
but the simple and distinctive cusps proposed here were not explored.

\begin{figure}[tb]
\centering
  \includegraphics[width=250pt]{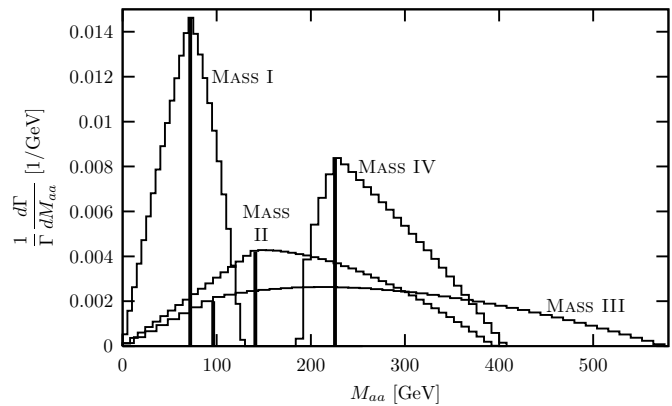}
  \caption{\label{fig:Mll}Normalized differential decay rates versus
the invariant mass $M_{aa}$
for various combinations of masses as given in Table \ref{table:massset}.
The vertical lines indicate the positions of the cusps in
each $M_{aa}$ distribution. }
\end{figure}

\begin{figure}
\centering
  \includegraphics[width=250pt]{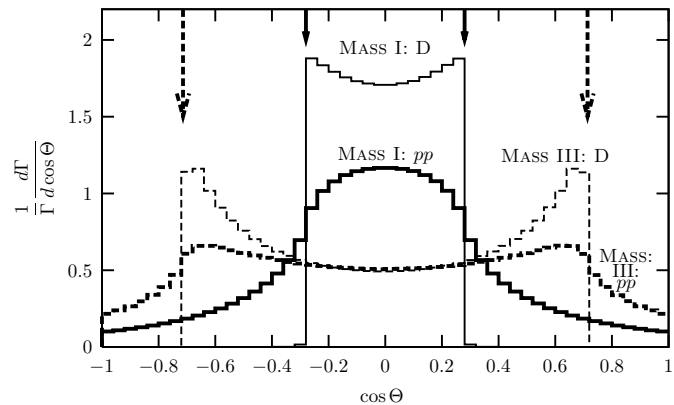}
  \caption{\label{fig:dGamdcosth}
Normalized differential decay rates versus
$\cos\Theta$ in the $D$-rest frame (thin curves) and in
the $pp$ lab frame with $\sqrt{s}=14\ \tev$ (thick curves). The parameters of
{\sc Mass I} and {\sc Mass III} are given in Table \ref{table:massset}. }
\end{figure}

For purposes of illustration, 
we explore  two kinematic distributions:
(\textit{i}) $M_{aa}$,
the invariant mass of $a_1$ and $a_2$, and
(\textit{ii}) $\cos\Theta$, cosine of the angle
between one of the two visible particles 
and the pair c.m.~moving direction in their c.m.~frame, as shown in Figs.~\ref{fig:Mll} and \ref{fig:dGamdcosth}
for various sets of parameter choice. 
The distributions have, in addition to the end points,  unique non-smooth
structures, the cusps (their positions are denoted by vertical lines).
While $M_{aa}$ and $\cos\Theta$ are not the only 
observables displaying the cusp structure,  these variables are advantageous 
since they do not involve the missing transverse energy.  

The appearance of the cusp can be understood intuitively as follows.  
We start with the flat distribution $d^2 \Gamma / d
\cos\theta_1 d \cos\theta_2$, where $\theta_1$ and $\theta_2$ are
the scattering angles of two visible particles
relative to the parent's boost direction
in their parent rest
frame,
and $\Gamma$ is the partial decay width 
of the particle $D$. Any
observable can be expressed as a function of $(\theta_1,\theta_2)$,
{\it e.g.}, $M_{aa}(\theta_1,\theta_2)$. Due to the ``antler'' decay
symmetry, $(\cos \theta_1, \cos \theta_2)$ and $(\cos \theta_2, \cos\theta_1)$ 
are kinematically equivalent.  
With this identification, the
result of projecting onto $M_{aa}$ is a folded space with three
distinctive points or apexes: The  lowest apex $(\pm1, \mp1)$ and
the highest apex $(1, 1)$ correspond to the endpoints $M_{aa}^{\rm
min}$ and $M_{aa}^{\rm max}$; while the third apex $(-1, -1)$ denotes
the position of the cusp, which occurs more frequently than the other two configurations. 

Cusps in the antler decay have a number of desirable features 
in determining the missing particle mass:
(\textit{i}) together with end-points, cusps can determine
both the masses of the intermediate particle $B$ and the missing particle $X$;
(\textit{ii})  looking for a cusp is statistically
advantageous since it has large (in most cases, maximum) event rate;
(\textit{iii})  there is no combinatoric complication due to its simple decay topology;
(\textit{iv}) spin correlations of the decay processes do not change the position of the cusps.
In the absence of backgrounds, 
cusps should be experimentally easy to 
identify due to their pointed features. 
In what follows, we will show that the cusp
position provides important information 
about the particle masses in the
decay process, which is complementary 
in many ways to the previously 
studied mass measurement methods.
We will also show that this information
is largely retained 
after the SM backgrounds as well as the detector simulation 
are included.

\section{Cusp and Edge in $M_{aa}$ Distribution }

We first only show the phase space distributions for on-shell particles.
It is convenient to use
the rapidities:
the rapidity $\eta$ of particle $B$
and the rapidity $\zeta$ of particle $a$
in the rest frames of their parents $D$ and $B$, respectively.
The rapidities $\eta$ and $\zeta$ are given by
\beq
\cosh \eta = \frac{m_D}{2 m_B}\equiv c_\eta, \ \
\cosh \zeta = \frac{m_B^2 - m_X^2 + m_a^2}{ 2 m_a m_B}\equiv c_\zeta .
\no
\eeq
\noindent
Here and henceforth we use a shorthand notation of
$c_{x}\equiv \cosh x$. Obtaining the rapidities would be
equivalent to measuring the masses $m_B$ and $m_X$.

\vskip 0.1cm
\noindent
{\textit{(1) $m_a=0$ case:}}
Consider $a$ to be massless first for simplicity.
One would naively expect the invariant mass to have an end-point
$M_{aa}^{\rm max} = m_D- 2m_X$. However, due to the on-shell constraint for the
particle $B$,
we find a different end-point:
\bea
\label{eq:Mmax:massless}
 M_{a a}^{\rm max }
 = m_B
\left(1-\frac{m_X^2}{m_B^2}\right)
e^{\eta}\,.
\eea
In addition, the $M_{aa}$ distribution has a cusp at
\bea
\label{eq:cusp:massless}
 M_{a a}^{\rm cusp }
= m_B
\left(1-\frac{m_X^2}{m_B^2}\right)
e^{-\eta} .
\eea
This is remarkable since the ratio
$M_{a a}^{\rm max} /M_{a a}^{\rm cusp} = e^{2\eta}$ is governed
by the initial decay $D \to BB$ and thus
gives $m_B$.
The product $M_{a a}^{\rm max}  M_{a a}^{\rm cusp}$ depends
on the secondary  decay $B \to a X$ and gives $m_X$.
Furthermore, $d \Gamma /d M_{aa}$ is
\begin{eqnarray}
\frac{d \Gm}{d M_{aa}}
\propto
\left\{
  \begin{array}{ll}
    2 \eta M_{aa}, & \hbox{if }
    0 \leq M_{aa} \leq M_{aa}^{\rm cusp}; \\[10pt]
    M_{aa} \ln \dfrac{M_{aa}^{\rm max}}{M_{aa}}, & \hbox{if }
M_{aa}^{\rm cusp} \leq M_{aa} \leq M_{aa}^{\rm max}.
  \end{array}
\right.
\label{eq:masslessinvmass}
\end{eqnarray}

\begin{table}
\centering
\begin{tabular}{|l|c|c|c|c|}
\hline
         & $m_D$ (GeV) & $m_B$ (GeV) & $m_a$ (GeV) & $m_X$ (GeV) \\
\hline
~{\sc Mass I}   & 1250 & 600 & 0 & 550 \\
~{\sc Mass II}  & 1000 & 440 & 0 & 300 \\
~{\sc Mass III} & 1000 & 350 & 0 & 200 \\
~{\sc Mass IV}  & 600  & 250 & $m_Z$ & 100 \\
\hline
\end{tabular}
\caption{\label{table:massset} Test mass spectrum sets
for mass measurements using kinematic cusp structure ($m_Z$
is the $Z$ boson mass). }
\end{table}

Figure \ref{fig:Mll} shows $d\Gamma / d M_{aa}$
for four sets of representative masses specified in Table \ref{table:massset}.
The choice of the parameters for \textsc{Mass I} is motivated by the $\zt$ decay in the
UED model\,\cite{ued}.
Since the two subsequent decays $\zt\to\lo\lo$ and $\lo\to \ell \gmo$
occur near the mass threshold,
\textsc{Mass I} is to be called the ``near threshold case''.
For comparison, we consider the mass parameters with sizable gap in \textsc{Mass III},
the ``large mass gap case.''
The visibility of the cusp depends on the ratio $M_{aa}^{\rm cusp}/M_{aa}^{\rm max}$.
As shown in Eq.~(\ref{eq:masslessinvmass}), the distribution
for $M_{aa}<M^{\rm cusp}_{aa}$ is linear,  while that after
$M^{\rm cusp}_{aa}$ is a concave curve with the maximum at $M_{aa}^{\rm max}/e$.
The cusp becomes a sharper peak
if $M_{aa}^{\rm cusp}\geq M_{aa}^{\rm max}/e$ (or equivalently $m_B>0.44 m_D$).
The parameters in \textsc{Mass} II are chosen
to represent this boundary case of $m_B \approx 0.44 m_D$.
The cusp structure is more pronounced for the near threshold case
(\textsc{Mass} I) than the large mass gap case 
(\textsc{Mass} III).

\vskip 0.1cm
\noindent
{\textit{(2) $m_a\neq 0$ case:}}
If the SM particle $a$ is massive (a $Z$ boson or a top
quark), we call it the ``massive case''.
The parameter choice
in \textsc{Mass IV}, motivated by the MSSM heavy 
Higgs boson decay
associated with two SM $Z$ bosons, is an example.
In this case, the analytic form of $d \Gamma /d M_{aa}$ is given by
three pieces (the explicit forms are not very illuminating and thus
not given here).
Its maximum  is
\begin{eqnarray}
M_{aa}^{\rm max}  = 2 m_a c_{\eta+\zeta}.
\label{eq:invmassmax}
\end{eqnarray}
The positions  of $M_{aa}^{\rm min}$ and $M_{aa}^{\rm cusp}$ are as follows,
depending on the relations of the two rapidities $\eta$ and $\zeta$:
\beq
\label{eq:Maa:min:cusp:massive}
\begin{array}{c|ccc}
      & ~~~ \eta <\zeta/2~~~ & ~~~ \zeta/2 < \eta < \zeta~~~&~~~ \zeta < \eta ~~~\\
\hline \\[-10pt]
M_{aa}^{\rm min}&  2 m_a & 2 m_a & 2 m_a{c}_{\eta-\zeta} \\
M_{aa}^{\rm cusp} & 2 m_a{c}_{\eta-\zeta} & 2 m_a{c}_\eta & 2 m_a{c}_\eta
\end{array}
\eeq
For all three regions in Eq.~(\ref{eq:Maa:min:cusp:massive}), the
$M_{aa}$ distribution shows a sharp cusp,
as illustrated by the \textsc{Mass} IV case in Fig.~\ref{fig:Mll}.
Note that the case $\zeta < \eta$ has different $M_{aa}^{\rm min}$,
rather than $2m_a$ as naively expected. It is
due to the enhanced boost of the two fast-moving
parent $B$'s. This shift helps to resolve the ambiguity among the three regions.
Still a two-fold ambiguity in the $\eta<\zeta/2$ and $\zeta/2<\eta<\zeta$ regions
remains since
we do not know whether the measured $M_{aa}^{\rm cusp}$ is $2 m_a {c}_{\eta-\zeta}$ or
$2 m_a {c}_\eta$.

We propose another independent observable to break this ambiguity,
$(\Delta |p_T^a|)_{\rm max}$,  which is the maximum of the difference
between the magnitudes of the two transverse momenta of $a_1$ and $a_2$:
\bea
(\Delta |p_T^a|)_{\rm max} &\equiv& {\rm max} \left(
\left|\vec{p}_{\,T}^{a_1} \right| -
\left|\vec{p}_{\,T}^{a_2} \right|
\right)
\\ \nonumber
&=& m_a \left[\sinh(\eta + \zeta) - \sinh \left| \eta-\zeta\right| \right],
\eea
which is invariant under longitudinal boosts.
Note that the two-fold ambiguity happens when $\eta<\zeta$.
With the observed $M_{aa}^{\rm max}$ and $M_{aa}^{\rm min}$,
$(\Delta |p_T^a|)_{\rm max} $ provides independent
information.

\section{Cusp in Angular Distribution}

We also have analyzed the distribution with respect to $\cos\Theta$,
where $\Theta$ is the angle of a visible particle, say $a_1$,
in the c.m.~frame of $a_1$ and $a_2$, with respect to their c.m.~moving direction.
The expression of $d\Gamma / d\cos \Theta$ in the $D$-rest
frame for $m_a=0$ is remarkably simple:
\begin{equation}
\frac{d\Gamma}{d \cos \Theta} \propto
\begin{cases}
\begin{array}{ll}
\sin^{-3} \Theta, &  \mbox{if $|\cos\Theta| \leq \tanh \eta$,} \\
0, & \mbox {otherwise.}
\end{array}
\end{cases}
\label{eq:costhdist}
\end{equation}
The distribution has a sharp end-point, another cusp,
with the highest event rate at the boundary:
\beq
\label{eq:costh:max}
\left| \cos \Theta \right|_{\rm max} = \tanh \eta
= \sqrt{1 - 4 m_B^2/m_D^2}\ .
\eeq
While this variable is unambiguous in the lab frame
at a lepton collider,  we cannot determine
the longitudinal motion of the particle $D$ 
in a hadron collider.
After convoluting with the parton distribution functions,
$d \Gamma/d\cos\Theta$  is smeared.
In Fig.~\ref{fig:dGamdcosth}, we compare the $\cos \Theta$ distribution
in the rest frame of $D$ (thin curves) with that in
the lab frame at the LHC (thick curves) for the near threshold case (\textsc{Mass I})
and the large mass gap case (\textsc{Mass III}).
We have assumed that $D$ is produced by direct $s$-channel
$gg$ or $q \bar{q}$ annihilation so that the longitudinal momentum
distribution of $D$ is obtained from the parton
distribution of the incident protons.

The convolution effects with the partons smear out the sharp $\cos\Theta$ cusp in the lab frame.
For the near threshold case (\textsc{Mass} I), the two sharp rises at both ends get much
less pronounced,
although  it is still possible to read the edge point off in the distribution.
For the large mass gap case (\textsc{Mass} III),
as the end point position approaches towards $\cos \Theta = \pm 1$,
the sharpness of two cusps maintains better.
It is interesting to note that the cusp in the $M_{aa}$ distribution and that in the
$\cos\Theta$ distribution provide complementary information for
determining $m_B^{}$.
The invariant mass distribution yields a better resolution
for the near threshold case,  while the angular distribution provides better one
for the large mass gap case.

To some extent, the $\cos\Theta$ distribution
in the $D$-rest frame may be obtained even in hadron collisions.
The smearing effects can be effectively modeled by the well-known
parton distribution functions in the large $x$ region.
Or the direct resonant decay of $D$ into SM particles
allows to extract the velocity distribution of $D$,
which can be used to recover $d \Gamma / d\cos\Theta$ in the $D$-rest frame.

\section{Discussion}

To this point, the discussions on the kinematic cusps
are rather theoretical, i.e., considering
only the kinematics
at parton level with 
perfect mass shell conditions, 
and ignoring the SM backgrounds 
as well as the detector simulation.
First we consider the effects of the matrix elements regarding
the spin correlations between the initial state and final state particles.
We have confirmed that, for the four processes in Eq.~(\ref{eq:model}),
including the full matrix elements does not change the
shape of the distributions.
The deviations from the phase space predictions become appreciable
when the fermions and vector bosons
(like the $\zt$ in UED)
have chiral couplings for both $D\to BB$ and $B\to aX$ decays.
Even in this extreme case,
the cusped peak remains at the same location and its
height is changed by about $ 2\%$.

\begin{figure}
\includegraphics[width=250pt]{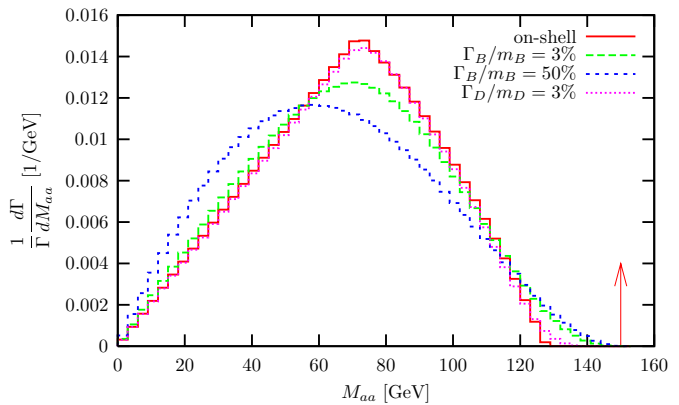}
\caption{\label{fig:GamB:Mll} The invariant mass distribution 
for the {\sc Mass I} case: $m_D^{} =1250\gev,\  m_B^{}=600\gev$,
with the finite decay widths $\Gm_D/m_D=3\%$ (dotted), and
$\Gamma_B/m_B = 3\%,\  50\%$ (long dashed and short dashed) respectively. }
\end{figure}

However, non-vanishing decay widths of 
the parent and intermediate particles can alter
the shape of the distributions. In Fig.~\ref{fig:GamB:Mll}, we show
the $M_{aa}$ distributions for different finite decay widths
$\Gamma_B$ and $\Gamma_D$ in the {\sc Mass I} spectrum. Finite
$\Gm_D$ has much milder effect than $\Gm_B$. 
With $\Gamma_B/m_B=3\%$, the finite decay width 
effect changes the distribution shape 
and the position of $M_{aa}^{\rm max}$, 
while essentially keeping the $M_{aa}^{\rm cusp}$ position. 
For illustration,
we also show a very broad case $\Gamma_B/m_B = 50\%$. This nearly off-shell
situation smears the triangular cusp shape considerably.
The momenta of the visible particles 
span all of the allowed
phase space given by $m_D$ and $m_X$, regardless of $m_B$.
The maximum of $M_{aa}$ approaches $m_D - 2 m_X$, denoted
by the arrow in Fig.~\ref{fig:GamB:Mll}.
The end point measurement with the known $m_D$
leads to the missing particle mass $m_X$, just like the direct neutrino mass
determination in tritium decays.
In the real scenarios listed in Eq.~(\ref{eq:model}),  the
intermediate particles ($\neu_2$, $\slt^\pm$, $\lo$ and $t_-$) typically have
decay widths smaller than one percent of their masses.
The $M_{aa}$ cusp shape remains intact.

We next explore  to what extent the SM backgrounds and
detector effects at the LHC would degrade the  sharp cusps.  
We consider a benchmark scenario in a SUSY model with 
an extra $U(1)$ gauge boson $Z'$.
In a minimal model where there is no mixing between $Z'$
and the SM gauge bosons,
the only relevant interactions of $Z'$ are with the SM
fermions and their SUSY partners, through the interaction
Lagrangian
$\Lg \supset g_1'Y_{f}\bar{f}\gm^\mu f Z'_\mu$.
Our signal process is
$Z' \to \tilde{\ell}^-  \tilde{\ell}^+ \to \ell^- \neuo  \ell^+\neuo$ 
with the mass parameters 
\begin{equation}
(m_{Z'}, m_{\tilde{\ell}}, m_{\chi^0_1}) = (1500, 730, 100)~{\rm GeV}.
\label{eq:masses}
\end{equation}
To manifest our signal, we take an optimistic scenario
where $g_1'=0.63$, $Y_{\rm quark}=Y_{\rm lepton}/2=1$, 
and all the sfermions except for
$\tilde{e}^\pm_{L,R}$ and $\tilde{\mu}^\pm_{L,R}$
are heavier than $m_{Z'}/2$.

The signal is two leptons with missing transverse energy.
The leading irreducible SM backgrounds are 
$W^+W^-\!/ ZZ \to \ell^+ \ell^-  \nu\bar \nu$.
The $t \bar{t}$ backgrounds decaying into
$b\bar{b} \ell^{+} \ell^{-} \nu \bar{\nu}$ can be dominant if not imposing very 
strong acceptance  cuts.
The SUSY backgrounds in this scenario 
are expected to be small:
(\textit{i}) the slepton pair production followed by 
the decay of $\tilde{\ell}^\pm\to \ell^\pm\tilde{\chi}^0_1$
is suppressed by the $P$-wave suppression and the heavy 
slepton mass, leading to the total cross section
of $\sim 0.7$ fb for the mass parameters in 
Eq.(\ref{eq:masses});
(\textit{ii}) another SUSY background of 
$pp \to \tilde{\chi}^+ \tilde{\chi}^-
\to \ell^+ \tilde{\nu} \ell^-\tilde{\nu}$
is extremely small because $m_{\tilde{\nu}}\gg m_{\tilde{\chi}_1^\pm}$
in this scenario;
(\textit{iii}) finally the rate of
$pp \to \tilde{\chi}^+ \tilde{\chi}^-
\to W^+W^-\tilde{\chi}^0_1\tilde{\chi}^0_1 $
followed by the leptonic decay of $W^\pm$ is also 
suppressed
in the Bino LSP scenario.
Therefore we ignore the SUSY background in what follows.

To suppress the SM top quark background, 
we veto the additional hard
jets in the kinematic region
%
\begin{eqnarray}
 E_{j} >
\left\{
  \begin{array}{ll}
    50~{\rm GeV} & \hbox{if } 3<|\eta_{j}|<5 , \\[10pt]
25~{\rm GeV}   & \hbox{if } |\eta_{j}| <3 ,
  \end{array}
\right.
\label{eq:ETcut}
\end{eqnarray}
where $\eta_{j}$ is the jet pseudo-rapidity. 
We have used \verb|MadGraph/MadEvent/PYTHIA|
\cite{MadGraphPYTHIA} for the event 
generation and \verb|PGS| \cite{PGS} for detector simulation.

\begin{figure}
\centering
  \includegraphics[width=250pt]{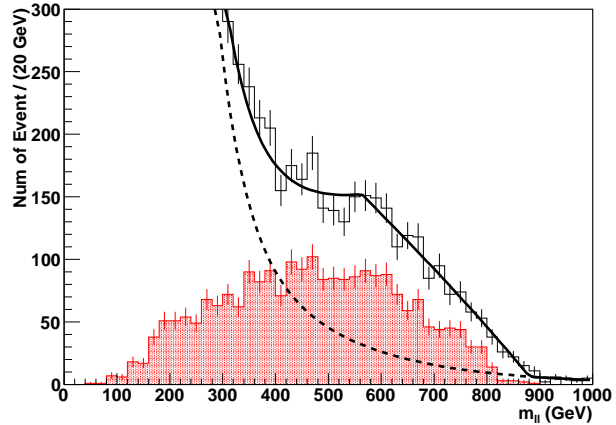}
  \caption{\label{fig:fullsim} Event distribution of the di-lepton invariant mass 
for the signal  $Z' \rightarrow \tilde{\ell}^+ \tilde{\ell}^- \rightarrow  \ell^{+} \ell^{-} \chi^0_1 \chi^0_1 $
as well as the SM backgrounds at the 14 TeV LHC and 100 fb$^{-1}$. 
The upper histogram with statistical error bars: signal plus background with
$\et > 50~{\rm GeV}$; 
the dotted curve for SM background only with $\et > 50~{\rm GeV}$.
Shaded histogram: signal with  $\et > 200~{\rm GeV}$. 
Mass parameters are given in Eq.~(\ref{eq:masses}). }
\end{figure}

In Fig.~\ref{fig:fullsim}, 
we show the lepton invariant mass distribution 
of the $Z'$ antler decay signal over the SM backgrounds
through $WW$, $WZ$, $ZZ$ and $t \bar{t}$ processes
at the LHC with c.m.~energy 14 TeV and luminosity 100 fb$^{-1}$.
The upper histogram with statistical error bars 
presents the signal plus backgrounds for
$\et > 50~{\rm GeV}$, and the dotted line is the SM background only. 
The SM background (especially $t\bar{t}$ ones) can 
still be substantial and comparable to the signal, although
the cusp feature and position can be clearly visible over the continuous background.
The shaded histogram shows the signal for
$\et > 200~{\rm GeV}$, and the SM background is essentially invisible with this
stringent missing energy cut. 
Although the signal can be made way above the SM background with
the characteristic  solitary triangular shape, 
the severe missing $\et$ cut alters the position of endpoint 
and smears the shape of the cusped peak. 
This is because the singly-produced heavy $Z'$ has little transverse motion, and the 
two missing particles are back-to-back for the cusp and the end-point configurations,
corresponding to low $\et$ situation. Some optimal treatment is needed with respect to the
selective kinematic cuts in order to effectively extract the missing particle mass. 


Once the signal is established with various cuts, 
we analyze the events with the low $\et$ cut (which faithfully respect the original 
cusp kinematics), and apply the known theoretical function in  Eq.~(\ref{eq:masslessinvmass}).
The best fit to the data curve leads to the reconstructed mass parameters as 
\begin{eqnarray}
&& m^{\rm recon}_{\tilde{\ell}} = 731.7^{+2.1}_{-5.2}~{\rm GeV}, \nonumber \\
&& m^{\rm recon}_{\chi_1^0} = 142.0^{+24.1}_{-39.7}~{\rm GeV}. \nonumber 
\end{eqnarray}
While the $\tilde{\ell}$ mass is determined with an impressive  accuracy, better than
a per-cent,  the neutralino mass is significantly shifted, with about a $50\%$ error. 
This uncertainty once again is mainly due to the distortion from $\et$ cut.

\section{Conclusions}

New techniques to measure the missing particle mass at the LHC are
proposed based on an antler decay topology ($D \to BB \to a X a X$),
with a final state of two visible particles $a$
and two missing particles $X$ via intermediate particles $B$.
We found a new type of kinematical singularity structure,
the cusps in the invariant mass distribution of two visible particles
and an angular distribution.
Its pure kinematical origin renders its cusp position and the distribution shape
nearly intact under the influence of the dynamical matrix elements.
Along with the end points of the distributions, the cusps can determine
the missing particle mass as well as the intermediate particle mass.
%
We demonstrated in a realistic example including the SM background analysis
and detector simulations that the signal is observable and the masses can be
determined to a reasonable accuracy. 

Our proposal relies on the observation of the antler decay, which 
are motivated in many new physics models. 
If such processes are seen in colliders, this method will 
provide a new way for mass measurement. 
We believe that this technique will be invaluable
for searches for new physics at the LHC and future lepton colliders,
as well as in any other processes with similar kinematics. 
The missing mass determination at colliders would undoubtedly shed light on 
the direct and indirect dark matter searches.

\begin{acknowledgments}
This work is supported in part by the U.S.
Department of Energy under grant No. DE-FG02-95ER40896.
The work of JS was supported by the
WCU program through the KOSEF funded by the MEST
(R31-2008-000-10057-0).
\end{acknowledgments}


%

\end{document}